\documentclass{aipproc}

\layoutstyle{6x9}

\newcommand\doingARLO[2][]{%
  \ifx\mmref\undefined #1\else #2\fi
}

\begin{document}

\title 
      {A Search for Supernova-Remnant Masers Toward Unidentified EGRET
	Sources}


\author{Z. Arzoumanian}{
  address={LHEA, NASA-GSFC, Code 662, Greenbelt, MD 20771},
  email={zaven@milklyway.gsfc.nasa.gov},
  thanks={NAS/NRC Research Associate}
}

\author{F. Yusef-Zadeh}{
  address={Dearborn Observatory, Northwestern University},
  email={zadeh@oort.astro.nwu.edu},
}

\author{T. J. W. Lazio}{
  address={Remote Sensing Division, Naval Research Laboratory},
  email={lazio@rsd.nrl.navy.mil}
}

\copyrightyear  {2001}

\begin{abstract}
Supernova remnants expanding into adjacent molecular clouds are believed
to be sites of cosmic ray acceleration and sources of energetic
gamma-rays. Under certain environmental conditions, such interactions
also give rise to unusual OH masers in which the 1720 MHz satellite line
dominates over the more common 1665/7 MHz emission. Motivated by the
apparent coincidence of a handful of EGRET sources with OH(1720 MHz)
maser-producing supernova remnants, we have carried out a search using
the Very Large Array for new OH(1720 MHz) masers within the error
regions of 11 unidentified EGRET sources at low Galactic latitude. While a
previously known maser associated with an HII region was serendipitously
detected, initial results indicate that no new masers were found down to 
a limiting flux of, typically, 50 mJy. We discuss the implications of
this result on the nature of the unidentified Galactic EGRET sources.
\end{abstract}


\maketitle

\section{Motivation}

A search for OH(1720 MHz) maser emission toward unidentified EGRET
sources is both theoretically and observationally well-motivated.

\begin{itemize}
\item Supernova remnant (SNR) shocks are believed to harbor sites of cosmic
ray acceleration and production of high-energy $\gamma$ rays: Fermi
acceleration may produce relativistic protons that interact with ambient
nuclei to create $\pi^0$, which decay into high-energy $\gamma$ rays.
A nearby molecular cloud can increase the density of target nuclei
(e.g., Aharonian et al.\ 1994).
OH maser emission is an unambiguous tracer of the type of
interactions likely to produce high-energy $\gamma$ rays
(Claussen et al.\ 1997; Wardle 1999).

\item The positions of $\gamma$-ray sources and SNRs on the sky are
correlated (e.g., Sturner \& Dermer 1995; 
Esposito et al.\ 1996),
especially for nearby remnants. This correlation has been attributed
to the presence of young, rotation-driven pulsars within the SNR,
but the alternative hypothesis (that the remnants themselves are the
$\gamma$-ray sources) has been largely unexplored observationally.

\item OH(1720 MHz) maser emission is detected from nearly two dozen Galactic
SNR adjacent to molecular clouds (e.g., Frail et al.\ 1996; Green 
et al.\ 1997; Yusef-Zadeh et al.\ 1999), including four remnants
associated with EGRET sources. These
sources have hard spectra and do not exhibit significant variability.

\item Perhaps coincidentally, SNRs found to contain OH masers and
hard low-latitude EGRET sources are both more prevalent in the inner 
Galaxy. Such a spectral disparity between inner- and outer-Galaxy EGRET
sources would be difficult to explain in a pulsar-origin model.
\end{itemize}

For these reasons, we chose the satellite line of the hydroxyl radical
(OH) at 1720.5 MHz as a tool to uncover remnants that may be obscured by
nearby or surrounding molecular clouds.

\section{OH(1720 MHz) Maser Emission}

OH(1720 MHz) masers have proven to be unique probes of C-type shocks,
the magnetic fields of SNRs behind the shock front, and gas dynamics
(Wardle 1999).
1720 MHz line emission is clearly evident in IC443, W28, W44
(Frail et al.\ 1996) and Sgr A East (Yusef-Zadeh et al.\ 1996),
all of which are coincident with EGRET sources. Three other SNRs
possibly associated with EGRET sources, CTA1, Monoceros, and
$\gamma$-Cygni, apparently do not contain OH masers---the
stringent criteria for producing such masers (an X-ray or cosmic
ray flux to dissociate H$_2$O formed in the shock to OH, and
molecular gas densities and temperatures appropriate for
collisional pumping of the OH; Wardle, Yusef-Zadeh \&
Geballe 1998) are not often met, even if particle acceleration
and $\gamma$-ray production are occuring. Indeed, Green et
al.\ 1997 find that only 10\% of the remnants they surveyed
contain OH masers, most of which belong to the morphological
class of radio shell, center-brightened thermal X-ray (``mixed morphology''; 
Rho \& Petre 1998) SNRs. By contrast, four of seven putative EGRET-SNR
associations exhibit OH maser emission.

\section{Target Selection}

Source selection was based on the Second EGRET Catalog, its Supplement,
and Lamb \& Macomb's (1997) catalog of GeV sources.
Telescope pointings were based on
source coordinates and error circles from the Third Catalog. We used
available corollary information for source spectra and variability
(Merck et al.\ 1996; McLaughlin et al.\ 1996).
Consistent with the properties of the existing (postulated) SNR associations,
two anti-center and nine inner-Galaxy EGRET sources were selected as
targets according to the following criteria:

\begin{itemize}
\item visible from the VLA: $\delta \ge -35^\circ$,
\item low Galactic latitude: $|b| \le 10^\circ$,
\item evidence of hard spectrum: $\alpha \le -2.0$ (for $F \propto
E^{-\alpha}$), or appearance in GeV source catalog.
\end{itemize}

\begin{table}[t]
\small
\begin{tabular}{rrrrccc}
\hline
\tablehead{1}{c}{b}{Source} & 
\tablehead{1}{c}{b}{$l$} & 
\tablehead{1}{c}{b}{$b$} & 
\tablehead{1}{c}{b}{$V$\tablenote{Variability index of McLaughlin et
al.\  1996. Values greater than 1 indicate $\sim 2 \sigma$ flux
variations on $\sim 1$ year timescales.}} & 
\tablehead{1}{c}{b}{$\alpha$/GeV\tablenote{Spectrum of $\gamma$-ray
flux, $F \propto E^{-\alpha}$; 
  ``GeV'' indicates $E_\gamma > 1$ GeV sources listed by Lamb \& Macomb (1997;
  ``?'' for their low-significance sources).
}} & 
\tablehead{2}{c}{b}{Notes\tablenote{
  ``C'': confused---source flux and significance may be uncertain.
  ``E'': may be extended.
}} \\
\hline
\tablehead{7}{c}{b}{OH(1720 MHz) Maser Search Targets.} \\
\hline
3EG J0459+3352 & 170.30 & $-$5.38 & 0.53 & 2.2 &C& 2EG J0506+3424\\
3EG J0634+0521 & 206.18 & $-$1.41 & 0.13 & 1.9 &C& 2EG J0635+0521\\
3EG J1734$-$3232 & 355.64 & 0.15 &       & GeV & & GEV J1732$-$3130\\
3EG J1809$-$2328 & 7.47 & $-$1.99 & 1.69 & 2.1/GeV &C & 2EG J1811$-$2338\\
3EG J1812$-$1316 & 16.70 & 2.39 & 3.05	& 2.3/GeV &C& 2EG J1813$-$1229\\
3EG J1823$-$1314 & 17.94 & 0.14 & 1.41	& 2.0/GeV &C& 2EG J1825$-$1307\\
3EG J1837$-$0606 & 25.86 & 0.40 &      & GeV &	 & GEV J1837$-$0611\\
3EG J1903+0550 & 39.52 & $-$0.05 &     & GeV &  & GEV J1907+0556\\
3EG J2021+3716 & 74.76 & 0.98 & 1.40	& 1.9/GeV &C& 2EG J2019+3719\\
3EG J2033+4118 & 75.58 & 0.33 &        & GeV &	 & GEV J2035+4210\\
3EG J2227+6122\tablenote{
In the time since our observations were made, X-ray and radio
counterparts to 3EG J2227+6122 have been proposed (Halpern et al.\
2001) that suggest a pulsar origin.
} & 106.53 & 3.18 & 0.34	& 2.1/GeV? &  & 2EG J2227+6122\\
\hline
\tablehead{7}{c}{b}{EGRET sources tentatively identified with SNRs.} \\
\hline
3EG J0010+7309  & 119.92 &10.54 & 1.49	& GeV &	 &  CTA1 \\
3EG J0617+2238  & 189.00 & 3.05 & 1.52 & 2.0 GeV & & IC443(OH) \\
GEV J0633+0645  & 204.83 &$-$0.96 &    & GeV &  & Monoceros \\
3EG J1746$-$2851  &  0.11  &$-$0.04 & 1.88 & GeV & & Sgr A E(OH) \\
3EG J1800$-$2338  &  6.25  &$-$0.18 & 0.05 & 1.9/GeV & C & W28(OH) \\
3EG J1856+0114  & 34.60  &$-$0.54 & 1.14 & GeV &   C & W44(OH) \\
2EG J2020+4017  & 78.05  & 2.08 & 0.83 & 2.1/GeV & C & $\gamma$-cygni \\
\hline
\end{tabular}
\caption{}

\end{table}

Telescope scheduling contraints and the sizes of the EGRET error circles
precluded a deep search of the full error region for each target of
interest. Our chosen observing strategy (integration time vs.\ number of
pointings) represents a compromise between sensitivity and coverage of
the error regions with $\sim 25'$-diameter FOV imaging. 

\begin{figure}
\caption{Contours of the 3EG likelihood test statistic
(Hartman et al.\ 1999) with two VLA fields-of-view superposed. The
asterisk indicates the position of a well-known HII region/OH maser.}
\includegraphics[width=3.75in]{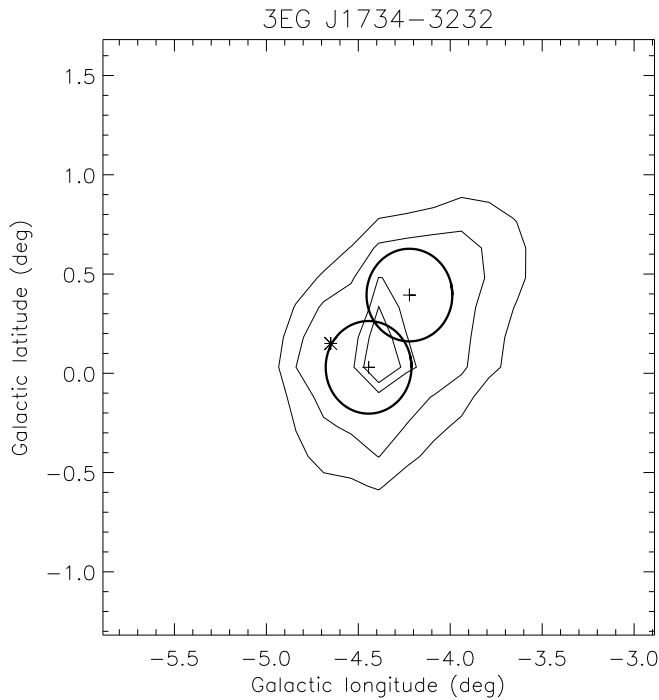}
\end{figure}

\section{Summary of Observations}

The 27 antennas of the NRAO Very Large Array were used in the CnD
configuration. Two IF pairs measured the right and left
circular polarizations simultaneously, with 128 channels for
each IF spanning a bandwidth of 1.5625 MHz. The IF pairs were 
centered at $v = \pm80$ km s$^{-1}$, yielding velocity coverage
of $\pm 216$ km s$^{-1}$ when the IFs were combined. 
Dwell times on each position were typically 10--12 minutes. 

The spectroscopic interferometer data were reduced using standard
procedures from the AIPS software package (Greisen 2000). A continuum
spectrum was fit and subtracted from each channel; the channels were
then imaged, and statistics for pixel intensities were formed.  We
searched for excesses in the normally-distributed noise over and above a
false-alarm probability of one pixel exceeding threshold, given the
number of channels and pixels (roughly 6$\sigma$). In addition to
searching for peaks in single channels, we used the SERCH prodecure
within AIPS to search for line profiles spread across 2, 3, and 4
channels, to increase sensitivity to weak line emission.

We observed the maser-producing remnant W28 as a test source:
its two brightest maser spots were readily detected in a very short 
(30 sec) integration.

\section{Results}

No new OH(1720 MHz) masers were detected in our survey, to a limiting
flux of roughly 50 mJy for the inner-Galaxy pointings, and 35 mJy for
the two anti-center pointings. Green et al.\ (1997) show (their Fig.\ 4)
that just two of the 119 OH masers from 17 SNRs have measured flux below
40 mJy, which suggests that our search is sensitive to maser emission
out to distances comparable to those of the $\sim 160$ remnants that 
have been surveyed for masers to date, i.e., better than half of all
cataloged SNRs, and for distances of at least several kpc.

We detected, serendipitously, line emission from the well-known hydroxyl
maser and HII region Maser 355.34+00.14 (in which the main OH lines at
1665 and 1667 MHz are not suppressed) as an unresolved source at
position RAJ 17:33:28.9, DecJ $-$32:47:49, near the edge of one FOV
(Fig.\  1).  The 1720 MHz line flux was $65\pm18$ mJy. Higher-resolution
observations (Forster \& Caswell 1999) distinguish a dozen maser spots with
1665 MHz line flux up to 17 Jy and LSR radial velocity $\sim 20$ km~s$^{-1}$. 
The HII region is unlikely to be associated with the EGRET source.

\section{Conclusions}

\looseness-1
The incompleteness of our survey, both in areal coverage of EGRET error
regions and, to a lesser extent, in sensitivity, makes it impossible to
draw firm conclusions about possible associations of unidentified EGRET
sources with supernova remnants driving cosmic-ray acceleration and
high-energy $\gamma$-ray emission. Nevertheless, and despite our present
null result, continuing multiwavelength follow-up observations of the
unidentified EGRET sources are bringing to light previously unknown
supernova remnants that are apparently associated with the $\gamma$-ray
sources---see, in particular, Combi, Romero \& Benaglia (1998) for 2EGS
J1703$-$6302 (beyond the VLA's southern limit), and Combi et al.\ (2001)
for three EGRET sources near $(\ell,b) = (6^\circ,-12^\circ)$. The
remnants they find are nearby, large, shell-like, and have low surface
brightness, and they do appear to abut against (low-density)
molecular clouds. If such
associations are borne out by future work and found to be common, the
apparent absence of OH(1720 MHz) maser emission suggested by our survey
may be attributable to the narrow range of environmental conditions under
which such emission is expected to arise (Green et al.\ 1997; Wardle
1999).

\begin{theacknowledgments}
The National Radio Astronomy Observatory is a facility of the National
Science Foundation operated under cooperative agreement by Associated
Universities, Inc.  This work was performed while ZA held a National
Research Council Research Associateship at NASA/GSFC. Basic research in
radio astronomy at the Naval Research Laboratory is supported by the
Office of Naval Research.
\end{theacknowledgments}



\end{document}